\documentclass[aps,prd,reprint,superscriptaddress,nofootinbib]{revtex4-2}
\usepackage{bbold}
\usepackage{amsmath,amssymb,amsfonts,float}
\usepackage{graphicx}
\usepackage{hyperref}
\usepackage{bm}
\usepackage{braket}
\usepackage{slashed}
\usepackage{subfig}
\usepackage[dvipsnames]{xcolor}
\usepackage{tikz}
\usetikzlibrary{calc, arrows.meta, bending, positioning, decorations.pathmorphing}

\graphicspath{ {./figures} }

\definecolor{TBlue}{HTML}{4967CD}
\definecolor{TRed}{HTML}{FF2E2E}
\definecolor{TOrange}{HTML}{FF7900}
\definecolor{TPurple}{HTML}{881EC4}
\definecolor{TBlueLight}{HTML}{DCECFF}
\definecolor{TRedLight}{HTML}{FFAAAA}
\definecolor{TOrangeLight}{HTML}{FFD6B6}
\definecolor{TPurpleLight}{HTML}{EAC4FF}
\definecolor{TGreyLight}{HTML}{E5E5E5}

\def\RadL{0.11cm}  % node radius for light operators
\def\RadH{0.16cm}  % node radius for heavy operators
\def\lU{0.16cm} % node size for unitary (square nodes)
\def\l{0.6cm}  % tensor leg length

\newcommand{\DrawLeg} {%
    \draw (P) -- ++(0:\l) coordinate (P);
}

\newcommand{\DrawNode}[4] {%
    \draw (P) -- ++(0:\l) coordinate (P);
    \coordinate (Q) at ($(P) + (#1,0)$);
    \coordinate (P) at ($(P) + (2*#1,0)$);
    \filldraw[fill=#3, draw=#4] (Q) circle (#1);
    \coordinate (#2) at (Q);
}

\newcommand{\DrawSquareNode}[3] {%
    \draw (P) -- ++(0:\l) coordinate (P);
    \coordinate (Q) at ($(P) + (\lU,0)$);
    \coordinate (P) at ($(P) + (2*\lU,0)$);
    \filldraw[fill=#2, draw=#3] ($(Q)+(-\lU, -\lU)$) rectangle ($(Q)+(\lU, \lU)$);
    \coordinate (#1) at (Q);
}

\newcommand{\DrawSquareNodeWithLegs}[3] {%
    \draw (P) -- ++(0:\l) coordinate (P);
    \coordinate (Q) at ($(P) + (\lU,0)$);
    \coordinate (P) at ($(P) + (2*\lU,0)$);
    \filldraw[fill=#2, draw=#3] ($(Q)+(-\lU, -\lU)$) rectangle ($(Q)+(\lU, \lU)$);
    \draw[draw=TBlue] ($(Q)+(-3pt,-\lU)$) -- ($(Q)+(-3pt, -\lU-0.6*\l)$);
    \draw[draw=TBlue] ($(Q)+(3pt,-\lU)$) -- ($(Q)+(3pt, -\lU-0.6*\l)$);
    \coordinate (#1) at (Q);
}

\newcommand{\DrawRhoNodeUp}[1] {%
    \draw (P) -- ++(0:\l) coordinate (P);
    \coordinate (Q) at ($(P) + (\RadH,0)$);
    \coordinate (P) at ($(P) + (2*\RadH,0)$);
    \filldraw[fill=TRedLight, draw=TRed] (Q) circle (\RadH);
    \node at (Q) {$\rho_R$};
    \coordinate (#1) at (Q);
}

\newcommand{\DrawRhoNodeDown}[1] {%
    \draw (P) -- ++(0:\l) coordinate (P);
    \coordinate (Q) at ($(P) + (\RadH,0)$);
    \coordinate (P) at ($(P) + (2*\RadH,0)$);
    \filldraw[fill=TRedLight, draw=TRed] (Q) circle (\RadH);
    \node at (Q) {$\rho_R$};
    \coordinate (#1) at (Q);
}

\newcommand{\DrawDanglingLegDown}[1] {%
    \draw[draw=TRed] (#1)+(0,-\RadH) -- ++(-90:\l);
}

\newcommand{\ConnectNodesAbove}[3] {
    \draw[draw=TRed] ($(#1)+(0,#3)$) edge[bend left=60] ($(#2)+(0,#3)$);
}

\newcommand{\ConnectNodesBelow}[3] {%
    \draw[draw=TRed] ($(#1)+(0,-#3)$) edge[bend left=-60] ($(#2)+(0,-#3)$);
}

\newcommand{\ConnectNodesAcross}[4] {
    \draw[draw=TRed] ($(#1)+(0,#3)$) edge[bend left=#4] ($(#2)+(0,-#3)$);
}

\newcommand{\InsertAR}[3] {%
  \draw[densely dashed] (#1) edge[bend left=#3]
  node[midway, draw, rectangle, solid, fill=white, inner sep=3pt] (TopBox) {$A_R$}(#2);
}

\newcommand{\InsertU}[2] {%
    \draw[fill=#2, draw=TBlue] ($(#1)+(-\lU, -\lU)$) rectangle ($(#1)+(\lU, \lU)$);
    \draw[draw=TBlue] ($(#1)+(-3pt,-\lU)$) -- ($(#1)+(-3pt, -\lU-0.8*\l)$);
    \draw[draw=TBlue] ($(#1)+(3pt,-\lU)$) -- ($(#1)+(3pt, -\lU-0.8*\l)$);
}

\newcommand{\ConnectSnaky}[5] {
  \coordinate (TmpC) at ($(#1)!1/2!(#2) + (0, #3)$);
  \draw[decorate, decoration={snake, amplitude=#5, segment length=4pt}] ($(#1)+(0,#4)$) .. controls ($(#1)!2/3!(TmpC)$) and ($(#2)!2/3!(TmpC)$) .. ($(#2)+(0,#4)$);
}

\newcommand{\DrawSnakyLeg}[5] {%
  \draw[decorate, decoration={snake, amplitude=#5, segment length=4pt}] (#1)+(0,#4) .. controls ($(#1)+(0,#3)$) .. ($(#2)+(0, #3)$);
}
\newcommand{\bu}{\mathbf{u}}

\newcommand{\bq}{\mathbf{q}}
\newcommand{\mc}{\mathcal}
\newcommand{\beq}{\begin{equation}}
\newcommand{\eeq}{\end{equation}}

\newcommand{\AB}[2][RoyalBlue]{\textcolor{#1}{(\textbf{AB:} #2)}}

\begin{document}

%\title{PETS, Python's Lunch and Magic}
\title{Stabilizer  complexity and the Python's lunch}
\author{Abhirup Bhattacharya}
\email{abhirup.bhattacharya@tifr.res.in}
\affiliation{Department of Theoretical Physics, Tata Institute of Fundamental Research, 1 Homi Bhabha Road, Mumbai 400005}

\author{Jatin Narde}
\email{jatin.narde@tifr.res.in}
\affiliation{Department of Theoretical Physics, Tata Institute of Fundamental Research, 1 Homi Bhabha Road, Mumbai 400005}

\author{Onkar Parrikar}
\email{parrikar@theory.tifr.res.in}
\affiliation{Department of Theoretical Physics, Tata Institute of Fundamental Research, 1 Homi Bhabha Road, Mumbai 400005}

\author{Suprakash Paul}
\email{suprakash.paul@tifr.res.in}
\affiliation{Department of Theoretical Physics, Tata Institute of Fundamental Research, 1 Homi Bhabha Road, Mumbai 400005}

\date{\today}

\begin{abstract}
In this note, we study the stabilizer complexity of the reduced density matrix corresponding to one side of a partially entangled thermal (PET) state with fixed energy boundary conditions in a holographic CFT. In particular, we study Wigner negativity, an operationally meaningful magic monotone which can be interpreted as the complexity of classically simulating any quantum circuit preparation of the reduced state on the subregion. Using assumptions on the pseudorandomness of the CFT spectrum and the heavy operator insertion, we observe that the Wigner negativity of the PET state relative to the microcanonical density matrix at the given energy is given by $\exp\left[\frac{1}{8G_N}(A_{\text{out}} - A_{\text{min}})\right]$, where $A_{\text{out}}$ is the area of the outer extremal surface, while $A_{\text{min}}$ is the area of the minimal extremal surface. Thus, the stabilizer complexity of the reduced density matrix on the boundary subregion is $O(1)$ in the absence of a python's lunch, but gets exponentially enhanced in the presence of a python's lunch in the bulk geometry. 

\end{abstract}

\maketitle

\section{Introduction}
Recent progress on the black hole information problem suggests that the interior of an evaporating black hole spacetime is encoded in the early radiation after Page time \cite{Penington:2019npb, Almheiri:2019psf, Almheiri:2019hni, Penington:2019kki, Almheiri:2019qdq}. From standard information theoretic considerations, this implies that the degrees of freedom in the black hole interior can be manipulated by acting on the radiation out at infinity, thus violating basic rules of quantum field theory on the black hole spacetime. Starting from the seminal work of Harlow and Hayden \cite{Harlow:2013tf}, several authors \cite{brown2020python, Kim:2020cds, Engelhardt:2021mue, Akers:2022qdl, Balasubramanian:2022fiy} have suggested that the semi-classical black hole geometry is nevertheless a good approximation for low-complexity observables, while operations which blatantly violate semi-classical causality are necessarily exponentially complex. In \cite{brown2020python}, this idea was geometrized and extended to general holographic states in the form of the ``python's lunch'' conjecture. Consider a holographic state where the bulk geometry has two extremal surfaces homologous to the same boundary subregion $R$. When the outer extremal surface is not the minimal area surface, then the portion of the entanglement wedge hidden behind the outer surface is called the python's lunch. The conjecture then states that reconstruction of bulk observables localized to this region as boundary operators on $R$ is exponentially complex. More precisely, the complexity of inverting the encoding map on $R$ is conjectured to be 
\beq \label{eq:PLC}
\mathcal{C} \sim \exp\left[\frac{1}{8G_N}(A_{\text{bulge}}- A_{\text{out}})\right],
\eeq 
where $A_{\text{out}}$ is the area of the outer extremal surface and $A_{\text{bulge}}$ is the area of a ``bulge'' surface, which in the time-reflection symmetric case we can think of the area of the maximal area surface between the two minimal surfaces. 

In this work, we study a different but related notion of complexity, namely the complexity of the subregion density matrix $\rho_R$, when the bulk geometry dual to the boundary state has a python's lunch. To keep things concrete, we will use a particular precisely defined and calculable notion of complexity called \emph{stabilizer complexity}. Given a computational basis on the underlying microscopic Hilbert space of $R$, the Gottesman-Knill theorem \cite{gottesman1998heisenberg, Aaronson:2004xuh} identifies a class of quantum operations called \emph{stabilizer operations}, which can be efficiently implemented in terms of a polynomial number of local gates \cite{Hostens:2005svl}, and efficiently simulated on a classical computer \cite{Mari_2012}. However, stabilizer operations are not universal, and the stabilizer complexity of a state is the amount of non-stabilizer or ``magic'' ancilla \cite{Bravyi_2005} needed to prepare the state. Intuitively, we can interpret this as the irreducible complexity of classically simulating any circuit preparation of the state \cite{PhysRevLett.115.070501}. More precisely, the stabilizer complexity can be quantified using the resource theory of stabilizer quantum computation \cite{Veitch_2012, veitch2014resource} which defines operationally meaningful measures called stabilizer or ``magic'' monotones.  

In \cite{Basu:2025uxw}, it was proposed that the stabilizer complexity of a boundary subregion gets exponentially enhanced when the bulk entanglement wedge of that subregion has a python's lunch. The purpose of this short note is to flesh out some details of this proposal in a particular class of states called Partially Entangled Thermal (PET) states \cite{Goel:2018ubv} with \emph{fixed-energy} boundary conditions. PET states are perhaps the simplest, prototypical examples of holographic states where the dual bulk geometry contains a python's lunch. Fixed energy boundary conditions will be very important in our discussion because it is only in this setting that we will be able to say something universal using assumptions on the chaotic nature of the CFT. We will be interested in a particular magic monotone, namely the \emph{Wigner negativity} of the reduced density matrix $\rho_R$ of, say, the right side of a PET state. For a finite dimensional Hilbert space, the discrete Wigner function \cite{WOOTTERS19871, Leonard, sphere, quantumcomp, Galois, Gross_2006, classicality} is a quasi-probability representation of quantum states on a discrete phase space. However, the discrete Wigner function is not a genuine probability distribution because it can take negative values, and remarkably, the amount of negativity $
\mathcal{N}(\rho_R)$ (see equation \eqref{negdef} for the definition) in the Wigner function turns out to be a magic monotone.  Unfortunately, this quantity has a universal (state-independent) divergence in the limit where the Hilbert space dimension diverges (for instance, the continuum limit on a lattice), so in order to avoid this issue, we will study the \emph{relative Wigner negativity}:
\beq 
\mathcal{N}(\rho_R | \rho_R^{(0)}) = \frac{\mathcal{N}(\rho_R)-1}{\mathcal{N}(\rho_R^{(0)})-1},
\eeq 
where $\rho_R^{(0)}$ will be taken to be the microcanonical ensemble, i.e., the maximally mixed state, in the same energy window. 

The essential point we wish to emphasize is this: assuming that the Hamiltonian for the holographic CFT restricted to a small enough energy window looks like a random matrix in the computational basis, it turns out that the calculation of the Wigner negativity reduces to the calculation for a Haar random state. Using known formulas in this case \cite{White:2020hgn} (see also \cite{Basu:2025mmm, Basu:2025uxw}), the negativity of the reduced density matrix $\rho_R$ of the PET state relative to the maximally mixed state at fixed energy $\rho_R^{(0)}$ (i.e., the microcanonical ensemble) is given by
\beq 
\mathcal{N}(\rho_R|\rho^{(0)}_R) \sim \exp\left[\frac{1}{8G_N}(A_{\text{out}}- A_{\text{min}})\right],
\eeq 
where $A_{\text{min}}$ is the area of the extremal surface with smallest area. Thus, the presence of a python's lunch in the geometry gives an exponential enhancement in the stabilizer complexity. Note that our formula for the stabilizer complexity differs from the python's lunch conjecture (see equation \eqref{eq:PLC}). This is not an immediate issue because the quantity we compute is not directly the complexity of bulk reconstruction; rather it is the complexity of the reduced state on the boundary subregion.   

The rest of this paper is organized as follows: in sec. \ref{sec:prelim}, we begin by reviewing some background material on PET states, stabilizer complexity etc. In sec. \ref{sec:calc} we present the calculation of Wigner negativity in fixed energy PET states using some pseudorandomness assumptions on the CFT spectrum and the operator insertion. We end with some concluding remarks in sec. \ref{sec:disc}.
\section{Preliminaries}
\label{sec:prelim}
\subsection{Partially entangled thermal states}
% It has been conjectured that 3D gravity is dual to an ensemble of large-$c$ CFTs with random OPE coefficients \cite{Collier:2019weq, Cotler:2020ugk, Chandra:2022bqq} and a sparse low energy spectrum \cite{Hartman:2014oaa} \AB{More papers to cite here}. The conjecture has been motivated by analogous results in two dimensions where it has been shown that JT (super)gravity, defined as a gravitational path integral, is dual to a random matrix theory \cite{saad2019jt, Stanford:2019vob}. In this section we will review the construction of \cite{Chandra_2023} which models the radial direction in spherically symmetric bulk geometries in terms of a random tensor network built out of ensemble averages of CFT structure constants. 

Consider two copies of a holographic quantum system with the Hilbert space $\mc{H}_L \otimes \mc{H}_R$. A conceptually clear way to construct such a doubled Hilbert space is to take $\mathcal{H}_R$ to be the dual Hilbert space $\mathcal{H}_L^\star$; then the doubled Hilbert space $\mc{H}_L \otimes \mc{H}_L^{\star}$ is the Hilbert space of operators acting on $\mathcal{H}_L$. Let $\Omega \in \mathcal{H}_L \otimes \mathcal{H}_R$ be the identity operator.  %Each copy of the CFT is endowed with a Hamiltonian and some distinguished set of ``simple'' operators $\phi_i$; we can think of these as operators that admit a good large-$N$ limit. 
Throughout this paper, we will study excited states of the form
%\begin{align} \label{eq:PETSnp}
%    |\Psi \rangle = \frac{1}{\sqrt{Z}}\, \widetilde{\mc{O}}_L\, \mc{O}_h\, \widetilde{\mc{O}}_R |\Omega\rangle,
%\end{align}
\begin{align} \label{eq:PETSnp}
    |\Psi \rangle = \frac{1}{\sqrt{Z}}\, e^{-\beta_L H_L}\, \mc{O}_h\, e^{-\beta_RH_R} |\Omega\rangle,
\end{align}
%where $\widetilde{\mc{O}}_L$ and $\widetilde{\mc{O}}_R$ are operators of the form\footnote{We have suppressed the spatial dependence of the operators for notational simplicity.} :
%\begin{align}
%    \widetilde{\mc{O}}_{L/R} &= e^{-\tau_1 H_{L/R}}\phi_1 e^{-\tau_2H_{L/R}} \phi_2 \cdots \phi_n e^{-\tau_n H_{L/R}}
%\end{align}
where $H_{L/R}$ is the Hamiltonian acting on the $L$ and $R$ factors respectively, and $\mc{O}_h$ is a spherically symmetric ``shell'' operator; in the above formula, we can think of $\mathcal{O}_h$ as acting either on the $L$ or the $R$ factor. On the CFT side, the only thing we will need to assume about $\mc{O}_h$ is that its matrix elements in the energy eigenstate basis are of the form
\beq \label{eq:ETH}
\langle E_i | \mc{O}_h|E_j\rangle = e^{-f(\overline{E},\omega)}\,R_{ij},
\eeq 
where $f$ is a smooth function of $\overline{E} = \frac{1}{2}(E_i+E_j)$ and $\omega = (E_i-E_j)$, and the coefficients $R_{ij}$ are \emph{pseudorandom}, which means that for a fixed theory they are fixed numbers, but to a good approximation look like they have been drawn randomly from an underlying probability distribution. The state \eqref{eq:PETSnp} now takes the form:
\begin{align} \label{eq:PETSnp2}
    |\Psi \rangle = \frac{1}{\sqrt{Z}}\,\sum_{i,j} e^{-\frac{1}{2}\beta_L E_{i} - \frac{1}{2}\beta_R E_{j}- f(\overline{E},\omega)}\, R_{ij}\, |E_{i}\rangle_L\otimes |E_j\rangle_R.
\end{align}

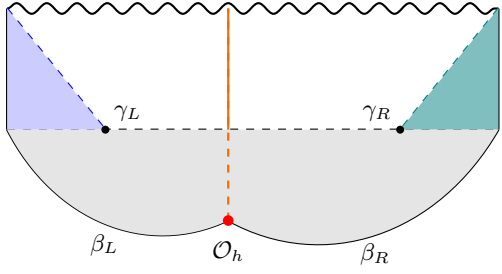
\begin{figure}
    \centering
    \begin{tikzpicture}[x=0.65cm, y=0.8 cm, >=latex]

% --- Coordinates ---
% Horizon at y=5, horizontal dashed line at y=3
% Left boundary x=0, right boundary x=10
% Intersection of 45° lines with horizontal line: γ_L=(2,3), γ_R=(8,3)
% O_H = (4.5, 1.5)

% --- Horizon (wavy line) ---
\draw[decoration={snake, amplitude=0.7mm, segment length=4mm}, decorate, thick] 
  (0,5) -- (10,5);

% --- Left and right vertical boundaries ---
\draw[thick] (0,5) -- (0,3);
\draw[thick] (10,5) -- (10,3);

% --- Horizontal dashed line (transition surface) ---
\draw[dashed, thick] (0,3) -- (10,3);

% --- 45° entanglement-wedge boundaries ---
\draw[dashed, thick, blue] (0,5) -- (2,3);   % left (blue)
\draw[dashed, thick, teal]  (10,5) -- (8,3);  % right (red)

% --- Euclidean sectors (convex from below) ---
% Left arc: smaller radius, from (0,3) to (4.5,1.5)
% Center = (3.166, 5), radius = 3.744, angles: 212.3° → 290.9° (counterclockwise)
\draw[thick] (0,3) arc (212.3:290.9:3.744);

% Right arc: larger radius, from (10,3) to (4.5,1.5)
% Radius = 4.5, angles from previous calculation
\draw[thick] (10,3) arc(-35.44:-114.06:4.5);

% --- Intersection points and labels ---
% --- Length labels for the arcs ---
\node at (1.98, 1.10) {\(\beta_L\)};   % midpoint of left arc
\node at (7.49, 0.95) {\(\beta_R\)};   % midpoint of right arc

% Fill left region
\fill[blue!20] (0,3) -- (0,5) -- (2,3) -- cycle;

% Fill right region
\fill[teal!50] (10,3) -- (10,5) -- (8,3) -- cycle;

\fill[TGreyLight] (0,3) -- (10,3) 
    arc(-35.44:-114.06:4.5) 
    -- (4.5,1.5) 
    arc(290.9:212.3:3.744) -- cycle;
    
% Infalling heavy particle
\draw[thick, red!20!orange] (4.5, 3)--(4.5, 5);
\draw[] [dashed, thick, red!20!orange] (4.5, 1.5)--(4.5, 5);

\fill (2,3) circle (1.5pt) node[above right] {\(\gamma_L\)};
\fill (8,3) circle (1.5pt) node[above left] {\(\gamma_R\)};
\fill[red] (4.5,1.5) circle (2pt);
\node[below] at (4.5,1.3) {\(\mathcal{O}_h\)};

\end{tikzpicture}
\caption{Euclidean path integral preparation of PET state on the $t=0$ slice. Lorentzian evolution of this state produces a geometry containing two black holes connected by a long wormhole.}
\label{fig:PETstate}

\end{figure}

States of the form equation \eqref{eq:PETSnp} are called \emph{partially entangled thermal} (PET) states, and just like the thermofield double state, they exhibit different phases depending on the values of $\beta_L$ and $\beta_R$. For not-too-large values of $\beta_{L,R}$, the bulk dual to such a state consists of two black holes connected by a long wormhole \cite{Goel:2018ubv} with two local extremal surfaces $\gamma_L$ and $\gamma_R$, as depicted in figure \ref{fig:PETstate}. The areas of $\gamma_L$ and $\gamma_R$ can be interpreted as the coarse-grained entropies of the $L$ and $R$ factors respectively, while the minimum of the two areas corresponds to the entanglement entropy. The causal complement of the two black hole exterior wedges constitutes a \emph{python's lunch} region \cite{brown2020python}, and contains within it the heavy bulk excitation dual to $\mathcal{O}_h$. In \cite{Chandra:2022fwi, Sasieta:2022ksu, Balasubramanian:2022gmo} it was shown that the function $f(\overline{E},\omega)$ in equation \eqref{eq:ETH} can be read off from the action of the bulk solution, and that if we take the coefficients $R_{ij}$ to be Gaussian random, then the ansatz in equation \eqref{eq:ETH} is consistent with bulk expectations from gravitational wormhole saddles.  

\subsection{Fixed energy boundary conditions}
% Between each pair of operators in equation \eqref{eq:PETSnp}, let us insert a resolution of identity in terms of energy eigenstates: 
%\begin{equation}
%    \mathbb{1} = \sum_{E_i} |E_i\rangle \langle E_i|.
%\end{equation}
%The state $|\Psi\rangle$ can now be written as
%\begin{align}\label{eq:TN1}
%    |\Psi \rangle = \frac{1}{\sqrt{Z}} \sum_{\text{all indices}} & (\widetilde{\mc{O}}_L)_{i_i; i_2} (\mc{O}_h)_{i_2;i_3} \nonumber \\ & \times (\widetilde{\mc{O}}_R)_{i_3; i_4}\, |E_{i_1} \rangle_L \otimes |E_{i_4} \rangle_R.
%\end{align}
%Each matrix element is a CFT 3-point function which takes the general form
%\begin{equation}
%    (\mc{O}(-\tau, \phi))_{\ell_i \lambda_i; \ell_j \lambda_j} = \mc{F}_{\lambda_i, \lambda_j}(-\tau, \phi | h_{\ell_i}, h_{\mc{O}}, h_{\ell_j})\, c_{\ell_i \mc{O} \ell_j}.
%\end{equation}
%The function $\mc{F}$ can be explicitly determined by acting on the 3-point function of primaries with a suitable combination of derivatives with respect to the location of the operator $\mc{O}$.
%Note that 
%\beq 
%(\mc{O}_h)_{\ell_2\lambda_2;\ell_3 \lambda_3} \propto c_{\ell_2,\ell_3,\mc{O}_h},
%\eeq
%is the CFT 3-point function between the primaries $\ell_2,\,\ell_3$ and $\mc{O}_h$ and is proportional to the above OPE coefficient; the other matrix elements involving $\widetilde{\mc{O}}_{L/R}$ are higher-point functions depending on the number of probe operator insertions. 
For the purposes of calculating correlation functions of simple operators, the sums over the left and right energies in equation \eqref{eq:PETSnp2} are dominated by semi-classical values of energies $(E_{L}^{\star}, E^\star_{R})$ which are determined via saddle-point equations in terms of the left and right inverse-temperatures $(\beta_L,\beta_R)$ and the mass of the shell \cite{Chandra_2023, Sasieta:2022ksu}. From the CFT point of view, the saddle-point equations arise from performing the energy integrals in a saddle-point approximation, while from the bulk point of view, these are simply bulk equations of motion. In this paper, we will find it convenient to truncate the sum over the energies to a small microcanonical window 
\beq \label{eq:micro}
\mc{H}_{\text{micro}}(E^\star)=\text{span}\Big\{|E_i\rangle \;\big|\; E^\star - \epsilon < E_i < E^\star + \epsilon\Big\},
\eeq
around the semi-classical value of the energy $E^\star$, where the size of the window should satisfy $\epsilon \ll E^\star$. From the bulk point of view, one can view this as implementing \emph{fixed energy}, i.e., Neumann boundary conditions, as opposed to the standard Dirichlet boundary conditions where the lengths of the Euclidean time segments are fixed at asymptotic infinity \cite{Brown:1992bq, Marolf:2018ldl, Chandrasekaran:2022eqq}. As we will see later, we will need to require that $\epsilon\sim E_{\text{RMT}}$, where $E_{\text{RMT}}$ is the scale of energy differences at which random matrix universality/ETH-like behavior is expected to emerge in chaotic quantum systems (see the discussion below equation \eqref{eq:MBBC}). This energy scale is expected to be of order $E_{\text{RMT}}\sim \frac{1}{\log(\frac{1}{G_N})}$ in holographic theories \cite{Gharibyan:2018jrp, YangStrings}.\footnote{By the uncertainty principle, such a microcanonical projection implies that the relative time shift mode has an uncertainty greater than $O(\log \frac{1}{G_N})$, which is large and so the state is not quite semiclassical. But the only aspect of the bulk geometry we will need to use is the classical areas of the extremal surfaces at leading order in $G_N$, which we do not expect to be affected by this uncertainty.}  %The dimension of the microcanonical window is governed by the Cardy entropy \cite{Chandra_2023}
The microcanonical entropies corresponding to the semi-classical energies $(E^\star_L,E^\star_R)$ at leading order in $\frac{1}{G_N}$ are given by the areas of the two extremal surfaces $\gamma_{L,R}$, i.e.
\begin{align}
    S(E^\star_L) & = \frac{A(\gamma_L)}{4 G_N}+\cdots \label{eq:EntropyRTL}, \\
    S(E^\star_R) &= \frac{A(\gamma_R)}{4 G_N} +\cdots \label{eq:EntropyRTR}.
\end{align}
Thus, one obtains the fixed-energy state,
%\begin{align}\label{eq:TN2}
%    |\Psi \rangle_{*} = \frac{1}{\sqrt{Z_\star}} \sum_{\substack{\text{all indices}\\ \in \{ \mathcal{H}_{E_i}^* \}}} & (\widetilde{\mc{O}}_L)_{i_i; i_2} (\mc{O}_h)_{i_2;i_3} \nonumber \\ & \times (\widetilde{\mc{O}}_R)_{i_3; i_4}\, |E_{i} \rangle_L \otimes |E_{j} \rangle_R.
%\end{align}
\begin{eqnarray}\label{eq:TN2}
    |\Psi \rangle_{*} &=&  \frac{1}{\sqrt{Z_\star}}\sum_{\substack{E_{i/j} \in \mathcal{H}_{\text{micro}}(E_{L/R}^\star)}} e^{-\frac{1}{2}\beta_L E_i - \frac{1}{2}\beta_R E_{j}}  \nonumber\\
    &\times & (\mathcal{O}_h)_{ij}\, |E_{i} \rangle_L \otimes |E_{j} \rangle_R.
   % &\sim &   \sum_{\substack{E_{i/j} \in \mathcal{H}_{\text{micro}}(E_{L/R}^\star)}}    R_{ij}\, |E_{i} \rangle_L \otimes |E_{j} \rangle_R.
\end{eqnarray}
Here, as indicated, we sum over only those states that lie within the truncated microcanonical windows, and $Z_\star$ is an appropriate normalization factor. %With that in mind, we can also approximate the state as
With this in mind, the heavy operator $\mathcal{O}_h$ can now be replaced with an effectively finite dimensional tensor as follows:
\begin{align}
    %\left(\tilde{\mathcal{O}}_{L}\right)_{i_1;i_2} &= 
   % \begin{tikzpicture}[baseline={(O)}]
       % \begin{scope}
       %     \coordinate (O) at (0, 0);
       %     \coordinate (P) at (O);
       %     \DrawNode{\RadL}{A}{TPurple}{TPurple}
       %     \DrawLeg
       %     \node[font=\small] at ($(A)+(0, 10pt)$) {$\widetilde{\mc{O}}_L$};
       %     \node at ($(O)+(-5pt, 0)$) {$i_1$};
       %     \node at ($(P)+(5pt, 0)$) {$i_2$};
       % \end{scope}
   % \end{tikzpicture} \\
    \left( \widehat{\mathcal{O}}_h \right)_{ij} &= \frac{e^{-\frac{1}{2}\beta_L E_i - \frac{1}{2}\beta_R E_{j}}}{\sqrt{Z_\star}}\times \,
    \begin{tikzpicture}[baseline={(O)}]
        \begin{scope}
            \coordinate (O) at (0, 0);
            \coordinate (P) at (O);
            \DrawNode{\RadH}{B}{TRed}{TRed}
            \DrawLeg
            \node[font=\small] at ($(B)+(0, 10pt)$) {$\mathcal{O}_h$};
            \node at ($(O)+(-5pt, 0)$) {$i$};
            \node at ($(P)+(5pt, 0)$) {$j$};
        \end{scope}
    \end{tikzpicture} %\\
   % \left(\tilde{\mathcal{O}}_{R}\right)_{i_3;i_4} &= 
   % \begin{tikzpicture}[baseline={(O)}]
   %     \begin{scope}
   %         \coordinate (O) at (0, 0);
   %         \coordinate (P) at (O);
   %         \DrawNode{\RadL}{C}{TOrange}{TOrange}
   %         \DrawLeg
   %         \node[font=\small] at ($(C)+(0, 10pt)$) {$\widetilde{\mc{O}}_R$};
   %         \node at ($(O)+(-5pt, 0)$) {$i_3$};
   %         \node at ($(P)+(5pt, 0)$) {$i_4$};
   %     \end{scope}
   % \end{tikzpicture},
\end{align}
%where it is henceforth understood that $\widetilde{O}_{L/R}$ are square matrices acting on $\mathcal{H}_{L/R}^\star$ respectively, and $O_h$ is 
where we are thinking of $\widehat{\mathcal{O}}_h$ as a rectangular matrix: 
$$ \widehat{\mathcal{O}}_h: \mathcal{H}_{\text{micro}}(E_R^\star) \to \mathcal{H}_{\text{micro}}(E_L^\star),$$ 
%As is conventional in tensor networks, contraction of two legs denotes summing over intermediate states, where recall that in our case, the sums are all restricted to the corresponding microcanonical windows, and hence are finite sums. 
and we have included the Boltzmann factors and the overall normalization constant in the definition above. In this language, the state $|\Psi \rangle_\star$ can be depicted pictorially as:
\begin{equation}\label{eq:RTN}
    |\Psi \rangle_{\star} =  \sum_{i, j}
    \begin{tikzpicture}[baseline={(O)}]
        \begin{scope}
            \coordinate (O) at (0, 0);
            \coordinate (P) at (O);
            %\DrawNode{\RadL}{A}{TPurple}{TPurple}
            \DrawNode{\RadH}{B}{TRed}{TRed}
            %\DrawNode{\RadL}{C}{TOrange}{TOrange}
            \DrawLeg
            % \DrawDanglingLegOut{B}
            \node[xshift=-5pt, yshift=1pt] at (O) {$i$};
            \node[xshift=5pt, yshift=0pt] at (P) {$j$};
            %\node[yshift=14pt, font=\small] at (A) {$\widetilde{\mathcal{O}}_{L}$};
            \node[yshift=14pt, font=\small] at (B) {$\widehat{\mathcal{O}}_h$};
            %\node[yshift=14pt, font=\small] at (C) {$\widetilde{\mathcal{O}}_{R}$};
        \end{scope}
    \end{tikzpicture}
    |E_{i}\rangle_L \otimes | E_{j} \rangle_R,
\end{equation}
where now the $i,j$ indices are implicitly summed over the respective microcanonical Hilbert spaces. %Likewise, the complex conjugates of the above tensors will be denoted by
In this presentation, we can interpret the state $\Psi_\star$ as a \emph{tensor network} \cite{Swingle:2012wq, Nozaki:2012zj, Miyaji:2016mxg, Bao:2018pvs, Caputa:2020fbc, Akers:2024wab}, with the bond dimensions of the left and right legs given by $e^{S(E^\star_L)}$ and $e^{S(E^\star_R)}$ respectively. If we further assume that $R_{ij}$ are Gaussian random variables, then the state becomes a \emph{random} tensor network \cite{Hayden:2016cfa, Chandra_2023, Akers:2021pvd}. 

\subsection{Stabilizer complexity}
In what follows, our goal will be to calculate the \emph{stabilizer complexity} of the above microcanonical truncation of the PET state. In the theory of quantum computation, there is a class of circuits called \emph{stabilizer circuits} which have the special property that they are efficiently classically simulable \cite{gottesman1998heisenberg, Aaronson:2004xuh, Mari_2012}.\footnote{For instance, any stabilizer/Clifford unitary can be synthesized from a polynomial number of elementary gates \cite{Hostens:2005svl}.} However, as one might expect, stabilizer circuits are not universal, and one needs additional ``magic'' gates to simulate arbitrary circuits \cite{Bravyi_2005}. In the resource theory of magic \cite{Veitch_2012, Veitch_2014}, stabilizer operations are regarded as free operations, while magic or non-stabilizerness is regarded as a resource (see \cite{White:2020zoz, Cao:2023mzo, cao2025gravitational, Basu:2025uxw, Malvimat:2026oqf, Bettaque:2026vpl, Benedetti:2026mfy} for some recent work on magic and stabilizer complexity in the gravity context). A physical interpretation of stabilizer complexity is that it measures the complexity of classical simulation using some quasi-probability distribution to sample classical trajectories \cite{PhysRevLett.115.070501}.\footnote{A word of caution: in the presence of a local tensor product structure, one can have special low-entanglement states which have large magic and yet are classically simulable on account of the low entanglement \cite{Vidal:2003pmm}.}    

In order to quantify magic, one defines operationally meaningful measures called \emph{magic monotones} -- these are measures which vanish on stabilizer states, and monotonically decrease (on average) under the action of stabilizer operations \cite{veitch2014resource}. In particular, this monotonicity property implies that if one starts with an initial state $\rho^{(0)}$ and implements a stabilizer protocol on it, then the probability $p_{\rho^{(0)}\to \rho}$ of obtaining the final output state $\rho$ is bounded above by:
\beq 
p_{\rho^{(0)}\to \rho} \leq \frac{\mathcal{M}(\rho^{(0)})}{\mathcal{M}(\rho)},
\eeq 
for any magic monotone $\mathcal{M}$. In this sense, magic monotones give quantitative bounds on the hardness of preparing $\rho$ from $\rho^{(0)}$ using stabilizer operations. 

In this paper, we will work with a particularly useful magic monotone called \emph{Wigner negativity}. For a finite dimensional quantum system comprising of $n$ qudits with $d$ being an odd prime,\footnote{When $D = d^n$ with $d$ being an odd prime, the Wigner function can be shown to be the unique quasi-probability distribution which ``geometrizes'' the action of the Clifford group in terms of symplectic affine transformations on phase space \cite{Gross_2006}.} there exists a natural quasi-probability representation called the \emph{discrete Wigner function} \cite{WOOTTERS19871, Leonard, sphere, quantumcomp, Galois, Gross_2006, classicality}. Let $\left\{|k\rangle\right\}_{k=0}^{d-1}$ be an ordered, orthonormal basis for each local qudit Hilbert space. The essential idea is to interpret this as the ``position'' basis, and correspondingly construct a ``phase space''. We define the \emph{discrete phase space} $\mathcal{P}_d$ as the lattice $\mathbb{Z}_d \times \mathbb{Z}_d$ of size $d^2$. %With respect to this basis, one defines a discrete version of the Heisenberg-Weyl (HW) operators:
%\beq 
%w(q,p) = e^{-\frac{2\pi i}{D}\frac{D+1}{2}qp}Z(p)X(q),
%\eeq 
%where $(q,p)\in \mathcal{P}$ and 
%\beq 
%Z(p)|q\rangle = e^{\frac{2\pi i p q}{D}}|q\rangle,\;\;\; X(q) |q'\rangle = |q'+q\rangle.
%\eeq
%HW operators are closed under multiplication:
%\beq \label{eq:multiplication}
%w(q,p)w(q',p') = e^{\frac{2\pi i}{D}\frac{(pq'-qp')}{2}}w(q+q',p+p').
%\eeq 
With respect to the chosen basis, we define a set of $d^2$ operators $A(q,p)$ for each qudit called phase-point operators, each labeled by a phase space point :
\begin{equation}\label{eq:A_Wooters}
A(q,p) = \sum_{k,\ell=0}^{d-1} \widehat{\delta}_{2q,k+l}e^{\frac{2\pi i}{D} (k-\ell)p }|k\rangle \langle \ell|,
\end{equation}
where the hatted Kronecker delta $\widehat{\delta}$ is the $\text{mod}(d)$ version, i.e., it is 1 when $(k+\ell) = 2q\,\text{mod}\,(d)$, and 0 otherwise. The phase point operators for the full system are then defined as
\beq 
A(\bu) = A(q_1,p_1)\otimes A(q_2,p_2)\otimes \cdots \otimes A(q_n,p_n),
\eeq 
where $\bu= (q_1,\cdots,q_n,p_1,\cdots,p_n)$. The discrete Wigner function for a density matrix $\rho$ is now defined as:
\beq 
W_{\rho}(\bu) = \frac{1}{D} \mathrm{Tr}\left(\rho A(\bu)\right),
\eeq 
 where $D = d^n$. The Wigner function attempts to represent the quantum state $\rho$ as a probability distribution in phase space, much like is the case in classical mechanics. Indeed, it satisfies many of the standard properties of a probability distribution: it is real, normalized and summing over $p$ (or $q$) gives the probability distribution in $q$ (respectively $p$). 
While these properties suggest that we should regard the Wigner function as a probability distribution in phase space, this interpretation fails for an important reason -- the Wigner function can take negative values at some points in phase space. Interestingly, it turns out that the \emph{sum negativity} of the Wigner function, defined as:
\begin{equation}
    \label{negdef}
    \mathcal{N}_s(\rho)= \frac{1}{2} \left(\sum_{\bu}|W_{\rho}(\bu)|-1\right),
\end{equation}
is a magic monotone, i.e., an operationally meaningful measure of non-stabilizerness \cite{Veitch_2014}. It is also convenient to define the quantity
\beq \label{eq:conv}
\mathcal{N}(\rho) := 1+ 2\mathcal{N}_s(\rho)= \sum_{\bu}|W_{\rho}(\bu)|,
\eeq 
which we will refer to simply as the Wigner negativity. For reasons that will become clear, we will actually consider the relative Wigner negativity 
\beq 
\mathcal{N}(\rho | \rho^{(0)}) := \frac{\mathcal{N}_s(\rho)}{\mathcal{N}_s(\rho^{(0)})} = \frac{\mathcal{N}(\rho)-1}{\mathcal{N}(\rho^{(0)})-1},
\eeq 
with respect to some other fiducial state $\rho^{(0)}$. As explained previously, the relative Wigner negativity gives an upper bound on the probability $p_{\rho^{(0)} \to \rho}$ using a stabilizer protocol:
\beq 
p_{\rho^{(0)} \to \rho} \leq \frac{1}{\mathcal{N}(\rho | \rho^{(0)})}.
\eeq 
Thus, the relative Wigner negativity gives a lower bound on the hardness of transforming $\rho^{(0)}$ to $\rho$ using a stabilizer protocol.
%So far we have discussed stabilizer complexity for states, but a similar notion can also be defined for quantum channels \cite{Mari_2012, Wang_2019}. For a quantum channel $\mathcal{E}(\rho) = \sum_m E_m\,\rho\,E_m^{\dagger}$ given in terms of its Krauss operator representation $\{E_m\}$, the Wigner function is defined as 
%\beq 
%W_{\mathcal{E}}(\bv|\bu)  = \frac{1}{D}\sum_m\mathrm{Tr}\left(A(\bv)\,E_m\,A(\bu)\,E_m^{\dagger}\right).
%\eeq 
%This can be thought of as a representation of the quantum channel as a quasi-stochastic matrix on phase space. Indeed, given an input state $\rho$ with the Wigner function $W_{\rho}(\bu)$, the Wigner function of the output state $\mathcal{E}(\rho)$ satisfies
%\beq 
%W_{\mathcal{E}(\rho)}(\bv) = \sum_{\bu} W_{\mathcal{E}}(\bv|\bu) \,W_{\rho}(\bu).
%\eeq 
%The negativity for a quantum channel is defined as
%\beq 
%\mathcal{N}(\mathcal{E}) = \text{max}_{\bu}\sum_{\bv}|W_{\mathcal{E}}(\bv|\bu)|.
%\eeq 
%Later we will be interested in computing the stabilizer complexity of bulk reconstruction using the Petz recovery channel, and we will use the above definition of Wigner negativity for this purpose. 

\section{Wigner negativity of the fixed energy PET state}
\label{sec:calc}
In this section, we will calculate the stabilizer complexity (as measured by Wigner negativity) for the reduced density matrix $\rho_R$ corresponding to the right subregion $R$ of the fixed-energy PET state $\Psi_\star$. The reduced density matrix can be pictorially represented as:
\begin{equation} \label{eq:rho_mc}
    \rho_{R} = 
    \begin{tikzpicture}[baseline={(O)}]
        \begin{scope}[scale=1.25]
            \coordinate (O) at (0, 0);
            \coordinate (P) at (O);
            %\DrawNode{\RadL}{A}{TOrangeLight}{TOrange}
            \DrawNode{\RadH}{B}{TRedLight}{TRed}
            %\DrawNode{\RadL}{C}{TPurpleLight}{TPurple}
            %\DrawNode{\RadL}{D}{TPurple}{TPurple}
            \DrawNode{\RadH}{E}{TRed}{TRed}
            %\DrawNode{\RadL}{F}{TOrange}{TOrange}
            \DrawLeg
            %\DrawDanglingLegDown{B}
            
            %\DrawDanglingLegDown{E}
            %\ConnectNodesBelow{B}{E}{\RadH}
            %\node[yshift=14pt, font=\small] at (A) {$\widetilde{\mc{O}}_R^\dagger$};
            \node[yshift=14pt, font=\small] at (B) {$\widehat{\mc{O}}_h^\dagger$};
            %\node[yshift=14pt, font=\small] at (C) {$\widetilde{\mc{O}}_L^\dagger$};
           % \node[yshift=14pt, font=\small] at (D) {$\widetilde{\mc{O}}_L$};
            \node[yshift=14pt, font=\small] at (E) {$\widehat{\mc{O}}_h$};
            %\node[yshift=14pt, font=\small] at (F) {$\widetilde{\mc{O}}_R$};
            \node[xshift=-7pt, font=\small] at (O) {};
            \node[xshift=7pt, font=\small] at (P) {};
        \end{scope}
    \end{tikzpicture}
\end{equation}

\noindent At this point, we will need to say a bit more about the microscopic nature of the Hamiltonian. We will work with the following setup: let $\mathcal{H}$ be the Hilbert space associated to some microscopic theory comprising of $n$ qudits (where $d$ is an odd prime) with some microscopic Hamiltonian, and consider the doubled Hilbert space $\mathcal{H}_L\otimes \mathcal{H}_R$. For instance, the microscopic system could be some generalized version of the Sachdev-Ye-Kitaev (SYK) model \cite{Sachdev_1993, Kitaev1, Kitaev2, Maldacena:2016hyu} comprising of $N$ qudits (with prime $d$), or some lattice regularization of a holographic CFT with a qudit at every site. We will assume that the microscopic system comes equipped with some natural choice of a local \emph{computational basis} for each qudit. Using this microscopic computational basis, we can construct the Wigner function for any density matrix on $R$. In particular, we will be interested in the Wigner function for the density matrix $\rho_R$ corresponding to the fixed-energy PET state:
\begin{equation}
W(\bu) = \frac{1}{D} \, \text{Tr}\!\left( \rho_R A_R(\bu) \right),
\end{equation}
where recall that $D=d^n$. Note that the density matrix $\rho_R$ has support on the $e^{S(E^\star_R)}$-dimensional subspace of the full Hilbert space. In order to employ the definition of the Wigner function written above, we will need to understand overlaps of the kind $\langle \bq | E_i\rangle$, where $|E_i\rangle$ is an energy eigenstate contained in the microcanonical Hilbert space $\mathcal{H}(E^\star_R)$ and $|\bq\rangle = |q_1,\cdots,q_n\rangle$ is an element of the computational basis. These overlaps are of course model-dependent, but in order to be able to say something universal, we will make the follow assumption: 
\\
\\
\noindent\textbf{Assumption 1}: \emph{For small enough $\epsilon$, the microcanonical subspace $\mathcal{H}(E^\star_R)$ looks like a random subspace relative to the computational basis.} 
\\
\\
To be more precise, the matrix elements
\beq \label{eq:MBBC}
U_{\bq, i} = \langle \bq| E_i\rangle,
\eeq 
can be well-approximated by matrix elements of a Haar random unitary. This assumption is essentially a many-body version of Berry's conjecture \cite{Berry:1977wpp, Lu:2017tbo}, a precursor of the eigenstate thermalization hypothesis \cite{Srednicki:1994mfb, DAlessio:2015qtq}. We expect this assumption to be satisfied by any chaotic system as long as the energy window $\epsilon$ in equation \eqref{eq:micro} is taken to be smaller than the energy scale $E_{\text{RMT}}$ at which the Hamiltonian shows universal random-matrix properties, in particular, randomness of energy eigenvectors with respect to a fixed computational basis. This energy scale should certainly be smaller than or equal to the Thouless energy \cite{DAlessio:2015qtq, Kos:2017zjh, Chan:2018dzt, Gharibyan:2018jrp, Thouless2, Winer:2020gdp, Chen:2023hra}. In holographic theories, we expect $E_{\text{RMT}}$ to be of order $\frac{1}{\log N}$ \cite{YangStrings}. Note that $e^{-S(E^{\star}_R)} \ll E_{\text{RMT}} \ll E^{\star}_R$. Some preliminary evidence in support of the above assumption in the context of the SYK
model is provided in Appendix \ref{appendix2}.

With this assumption, the reduced density matrix 
$ \rho_R = \sum_{i,j \in \mathcal{H}(E^\star_R)} \rho_{ij} |E_i\rangle\langle E_j|$
can be written in terms of the computational basis as 
\beq 
\rho_R = \sum_{\bq,\bq'} \sum_{i,j \in \mathcal{H}(E_R^\star)} U_{\bq, i} \rho_{ij} U^*_{\bq',j}\,|\bq\rangle\langle \bq'|.
\eeq 
%$\rho_{R,\mathrm{mc}}$ with zeros outside its support,
%\begin{equation*}
%\rho_{R, \text{full}}
%=
%\begin{pmatrix}
%\rho_{R,\mathrm{mc}} & 0 \\
%0 & 0
%\end{pmatrix},
%\end{equation*}
%One justification for the above assumption is that it is expected that in the black hole sector, the Hamiltonian looks like a random matrix with respect to any computationally simple basis. The key point is that, due to the randomness of the boundary Hamiltonian, the subspace associated with the microcanonical window is itself not fixed relative to a particular choice of basis of the full $D$-dimensional Hilbert space. We can model this randomness by conjugating $\rho_{R,\text{full}}$ with Haar-random unitaries,
%\begin{equation}
%\rho_{R}
%= U_{D\times D}
%\begin{pmatrix}
%\rho_{R,\mathrm{mc}} & 0 \\
%0 & 0
%\end{pmatrix} U_{D\times D}^\dagger
%\end{equation}
%In our subsequent discussion, we will be using this form of the density matrix and average over the Haar-random unitaries using the standard methods. Of course, we will average over the matrix elements of the heavy operator as well.

Using the above formula, we can diagrammatically write the Wigner function as: 
\begin{equation} \label{eq:wigner}
    W = \frac{1}{D } \hspace{0.2cm}
    \begin{tikzpicture}[baseline={(O)}]
        \begin{scope}[scale=0.75]
            \def\l{1cm};
            \def\RadH{0.28cm}'
            \def\lU{0.28cm};
            \coordinate (O) at (0, 0);
            \coordinate (P) at (O);
            %\DrawNode{\RadL}{A}{TOrangeLight}{TOrange}
            \DrawNode{\RadH}{B}{TRedLight}{TRed}
            %\DrawNode{\RadL}{C}{TPurpleLight}{TPurple}
            %\DrawNode{\RadL}{D}{TPurple}{TPurple}
            \DrawNode{\RadH}{E}{TRed}{TRed}
            %\DrawNode{\RadL}{F}{TOrange}{TOrange}
            \DrawLeg
           % \DrawDanglingLegDown{B}
            %\DrawDanglingLegDown{E}
            \InsertAR{O}{P}{45}
            \draw[fill=TBlueLight, draw=TBlue] ($(O)+(-\RadH, -\RadH)$) rectangle ($(O)+(\RadH, \RadH)$);
            \InsertU{O}{TBlue}
            \InsertU{P}{TBlueLight}
            \node at (O) {\textcolor{white}{$U$}};
            \node at (P) {$U^\dagger$};
        \end{scope}
    \end{tikzpicture}
\end{equation} \\

\noindent Given that $U_{\bq,i}$ can be approximated as the matrix elements of a Haar random unitary, we can now calculate the negativity of the above Wigner function by averaging over Haar random unitaries $U$. This calculation is essentially equivalent to the calculation of the Wigner negativity of Haar random states \cite{White:2020hgn, Basu:2025mmm, Basu:2025uxw}, but we will briefly explain the details of the calculation below for completeness. We will also argue that the fluctuations in the negativity relative to the averaged value are suppressed in $\frac{1}{D}$.

\subsection{Replica trick method}

To begin with, we wish to evaluate the averaged negativity:
\begin{equation}
    \mathcal{N}_{\text{avg}}=\sum_{\bu} \langle \; |W(\bu)| \; \rangle_{U},
\end{equation}
where the symbol $\langle \cdots\rangle_U$ denotes averaging over the Haar random unitary $U$. Note that the averaged Wigner negativity need not be the same as the negativity of the averaged Wigner function; it is important to treat the absolute value in the definition of the negativity with care. To evaluate the average of the absolute value of the Wigner function, we employ the \emph{replica trick}: 
\begin{equation}
    \mathcal{N}_{\text{avg}}=\sum_{\bu} \; \lim_{n\to\frac{1}{2}} \langle W^{2n}(\bu) \rangle_{U}.
\end{equation}
Our task is therefore to first compute $\langle W^{2n} \rangle_{U}$ and subsequently perform an analytic continuation to $n = \tfrac{1}{2}$ in order to obtain the average of the absolute value. There is a more systematic way to compute the negativity without resporting to analytic continuation; this method will be explained in Appendix \ref{appendix1}.

\subsubsection{Haar averaging}

To begin, we consider $2n$ copies of the diagram shown in \ref{eq:wigner}. For the purposes of this discussion, the detailed structure of $\rho_R$ in equation \eqref{eq:wigner} is not important and may be represented schematically by a blob. Hence, pictorially, we have:

\begin{widetext}
    \begin{equation}
    W^{2n}=\frac{1}{D^{2n}} \;\;\;
    \begin{tikzpicture}[baseline={(O)}]
        \begin{scope}[scale=0.75]
            \def\l{1cm};
            \def\RadH{0.3cm}'
            \def\lU{0.28cm};
            \coordinate (O) at (0,0);
            \coordinate (P) at (O);
            \DrawSquareNodeWithLegs{A}{TBlue}{TBlue};
            \DrawRhoNodeDown{B}
            \DrawSquareNodeWithLegs{C}{TBlueLight}{TBlue};
            \DrawLeg;
            \InsertAR{O}{P}{45};
            
            \node at (A) {\textcolor{white}{$U$}};
            \node at (C) {$U^\dagger$};
        \end{scope}
    \end{tikzpicture}
    \times \;
    \begin{tikzpicture}[baseline={(O)}]
        \begin{scope}[scale=0.75]
            \def\l{1cm};
            \def\RadH{0.3cm}'
            \def\lU{0.28cm};
            \coordinate (O) at (0,0);
            \coordinate (P) at (O);
            \DrawSquareNodeWithLegs{A}{TBlue}{TBlue};
            \DrawRhoNodeDown{B}
            \DrawSquareNodeWithLegs{C}{TBlueLight}{TBlue};D
            \DrawLeg;
            \InsertAR{O}{P}{45};
            
            \node at (A) {\textcolor{white}{$U$}};
            \node at (C) {$U^\dagger$};
        \end{scope}
    \end{tikzpicture}
    \times \cdots \cdots \;\times \;
    \begin{tikzpicture}[baseline={(O)}]
        \begin{scope}[scale=0.75]
            \def\l{1cm};
            \def\RadH{0.3cm}'
            \def\lU{0.28cm};
            \coordinate (O) at (0,0);
            \coordinate (P) at (O);
            \DrawSquareNodeWithLegs{A}{TBlue}{TBlue};
            \DrawRhoNodeDown{B}
            \DrawSquareNodeWithLegs{C}{TBlueLight}{TBlue};D
            \DrawLeg;
            \InsertAR{O}{P}{45};
            
            \node at (A) {\textcolor{white}{$U$}};
            \node at (C) {$U^\dagger$};
        \end{scope}
    \end{tikzpicture}
\end{equation}
We now perform the Haar average using the formula:
\begin{equation} \label{eq:Haargeneral}
    \frac{1}{{\text{Vol}_{U(D)}}} \int dU\, U_{\bq_1, j_i} \cdots U_{\bq_{2n}, j_{2n}} U^*_{\bq_1', j_1'} \cdots U^*_{\bq'_{2n},j'_{2n}} = \sum_{\sigma, \tau \in S_{2n}} \text{Wg}(\sigma \tau^{-1}, D) \delta_{\bq_1 \bq'_{\sigma(1)}} \cdots \delta_{\bq_{2n} \bq'_{\sigma(2n)}} \delta_{j_1 j'_{\tau(1)}} \cdots \delta_{j_n j'_{\tau_{2n}}}.
\end{equation}
% \begin{multline}
%     \frac{1}{\text{Vol}_{U(D)}}\int dU \; U_{i_1 j_1} U_{i_2 j_2}....U_{i_{2n} j_{2n}}\; U^*_{i_1'j_1'} U^*_{i_2' j_2'}....U^*_{i'_{2n}j'_{2n}}=\;\\\sum_{\sigma, \tau\; \in S_{2n}} \text{Wg}(\sigma\tau^{-1},D)\;\; \delta_{i_1 i'_{\sigma(1)}} \delta_{i_2 i'_{\sigma(2)}}...\delta_{i_n i'_{\sigma(2n)}}
%     \;\\ \delta_{j_1 j'_{\tau(1)}} \delta_{j_2 j'_{\tau(2)}}...\delta_{j_n j'_{\tau(2n)}} .\label{eq:Haargeneral}
% \end{multline}
Using the fact that the Weingarten functions factorize in the large $D$ limit (see the discussion below eq. C5 of \cite{Basu:2025uxw}), one can reorganize the series into a form in which any generic term in the expansion can be written as a product of $n_1$ one-body contractions, $n_2$ two-body contractions, and so on, such that $\sum_{p=1}^{2n} p \, n_p = 2n$. Here, a $p$-body contraction is defined as follows:
we take $p$ copies of the Wigner function in \eqref{eq:wigner} and contract all occurrences of $U$ and $U^\dagger$ among these copies in such a way that the resulting diagram is fully connected. Of course, there will be a number of such diagrams. One of these diagrams is associated with the Weingarten factor $\text{Wg}(1^p,D)$ which corresponds to the contraction type $\left( \langle UU^\dagger \rangle \right)^p$, with the additional requirement that all $p$ copies are connected into a single fully connected diagram. The remaining diagrams necessarily involve one or more higher-point connected correlators of $U$'s, such as $\langle UU^\dagger UU^\dagger\rangle_{U,\;\text{conc}}$ and their higher-order generalizations. The corresponding Weingarten factors of those diagrams are suppressed at large $D$ relative to $Wg(1^p,D)$. However, we denote the sum of all of these by $\mc{D}_p$. As an illustration of the above discussion, we show a few examples below. 
\begin{equation}
    \mc{D}_1 =
    \begin{tikzpicture}[baseline={(O)}]
        \begin{scope}[scale=0.8]
            \def\l{1cm};
            \def\RadH{0.3cm}'
            \def\lU{0.28cm};
            \coordinate (O) at (0,0);
            \coordinate (P) at (O);
            \DrawSquareNode{A}{TBlue}{TBlue};
            \DrawRhoNodeDown{B}
            \DrawSquareNode{C}{TBlueLight}{TBlue};D
            \DrawLeg;
            \InsertAR{O}{P}{45};
            \draw[draw=TBlue] ($(A)+(3pt,-\lU)$) edge[bend left=-60] ($(C)+(-3pt,-\lU)$);
            \draw[draw=TBlue] ($(A)+(-3pt,-\lU)$) edge[bend left=-70] ($(C)+(3pt,-\lU)$);
            \node at (A) {\textcolor{white}{$U$}};
            \node at (C) {$U^\dagger$};
        \end{scope}
    \end{tikzpicture}
\end{equation}

\begin{equation}
    \mc{D}_2 =
    \begin{tikzpicture}[baseline={(O)}]
        \begin{scope}[scale=0.75]
            \def\l{1cm};
            \def\RadH{0.35cm};
            \def\lU{0.3cm};
            \coordinate (O) at (0,0);
            \coordinate (O1) at (0, 1.2cm);
            \coordinate (O2) at (0,-1.2cm);
            \coordinate (P) at (O1);
            \DrawSquareNode{A}{TBlue}{TBlue};
            \DrawRhoNodeDown{B}
            \DrawSquareNode{C}{TBlueLight}{TBlue};
            \DrawLeg;
            \coordinate (P1) at (P);
            \InsertAR{O1}{P1}{45};
            \coordinate (P) at (O2);
            \DrawSquareNode{D}{TBlueLight}{TBlue};
            \DrawRhoNodeUp{E}
            \DrawSquareNode{F}{TBlue}{TBlue};
            \DrawLeg;
            \coordinate (P2) at (P);
            \InsertAR{O2}{P2}{-45};
            \draw[draw=TBlue] ($(A)+(3pt,-\lU)$) edge[bend left=0] ($(D)+(3pt,\lU)$);
            \draw[draw=TBlue] ($(A)+(-3pt,-\lU)$) edge[bend left=0] ($(D)+(-3pt,\lU)$);
            \draw[draw=TBlue] ($(C)+(3pt,-\lU)$) edge[bend left=0] ($(F)+(3pt,\lU)$);
            \draw[draw=TBlue] ($(C)+(-3pt,-\lU)$) edge[bend left=0] ($(F)+(-3pt,\lU)$);
            \node at (A) {\textcolor{white}{$U$}}; 
            \node at (C) {$U^\dagger$};
            \node at (D) {$U^\dagger$};
            \node at (F) {\textcolor{white}{$U$}};
        \end{scope}
    \end{tikzpicture}
    +
    \begin{tikzpicture}[baseline={(O)}]
        \begin{scope}[scale=0.75]
            \def\l{1cm};
            \def\RadH{0.35cm};
            \def\lU{0.3cm};
            \coordinate (O) at (0,0);
            \coordinate (O1) at (0, 1.2cm);
            \coordinate (O2) at (0,-1.2cm);
            \coordinate (P) at (O1);
            \DrawSquareNode{A}{TBlue}{TBlue};
            \DrawRhoNodeDown{B};
            \DrawSquareNode{C}{TBlueLight}{TBlue};
            \DrawLeg;
            \coordinate (P1) at (P);
            \InsertAR{O1}{P1}{45};
            \coordinate (P) at (O2);
            \DrawSquareNode{D}{TBlueLight}{TBlue};
            \DrawRhoNodeUp{E};
            \DrawSquareNode{F}{TBlue}{TBlue};
            \DrawLeg;
            \coordinate (P2) at (P);
            \InsertAR{O2}{P2}{-45};
            \draw[draw=TBlue] ($(A)+(-3pt,-\lU)$) edge[bend left=0] ($(D)+(-3pt,\lU)$);
            \draw[draw=TBlue] ($(A)+(3pt,-\lU)$) edge[bend left=-50] ($(C)+(-3pt,-\lU)$);
            \draw[draw=TBlue] ($(D)+(3pt,\lU)$) edge[bend left=50] ($(F)+(-3pt,\lU)$);
            \draw[draw=TBlue] ($(C)+(3pt,-\lU)$) edge[bend left=0] ($(F)+(3pt,\lU)$);
            \node at (A) {\textcolor{white}{$U$}}; 
            \node at (C) {$U^\dagger$};
            \node at (D) {$U^\dagger$};
            \node at (F) {\textcolor{white}{$U$}};
        \end{scope}
    \end{tikzpicture}\\
    +
    \begin{tikzpicture}[baseline={(O)}]
        \begin{scope}[scale=0.75]
            \def\l{1cm};
            \def\RadH{0.35cm};
            \def\lU{0.3cm};
            \coordinate (O) at (0,0);
            \coordinate (O1) at (0, 1.2cm);
            \coordinate (O2) at (0,-1.2cm);
            \coordinate (P) at (O1);
            \DrawSquareNode{A}{TBlue}{TBlue};
            \DrawRhoNodeDown{B}
            \DrawSquareNode{C}{TBlueLight}{TBlue};
            \DrawLeg;
            \coordinate (P1) at (P);
            \InsertAR{O1}{P1}{45};
            \coordinate (P) at (O2);
            \DrawSquareNode{D}{TBlueLight}{TBlue};
            \DrawRhoNodeUp{E};
            \DrawSquareNode{F}{TBlue}{TBlue};
            \DrawLeg;
            \coordinate (P2) at (P);
            \InsertAR{O2}{P2}{-45};
            \draw[draw=TBlue] ($(A)+(-3pt,-\lU)$) edge[bend left=-50] ($(C)+(3pt,-\lU)$);
            \draw[draw=TBlue] ($(A)+(3pt,-\lU)$) edge[bend left=0] ($(D)+(3pt,\lU)$);
            \draw[draw=TBlue] ($(D)+(-3pt,\lU)$) edge[bend left=50] ($(F)+(3pt,\lU)$);
            \draw[draw=TBlue] ($(C)+(-3pt,-\lU)$) edge[bend left=0] ($(F)+(-3pt,\lU)$);
            \node at (A) {\textcolor{white}{$U$}}; 
            \node at (C) {$U^\dagger$};
            \node at (D) {$U^\dagger$};
            \node at (F) {\textcolor{white}{$U$}};
        \end{scope}
    \end{tikzpicture}
\end{equation}
\end{widetext}
So, we can write,
\begin{equation}\label{eq:W2n_haar}
    \langle W^{2n}\rangle_U=\frac{1}{\left( D \right)^{2n}}\sum_{\{n_p \}}' \mathcal{S}\left(\{n_p\}\right) \prod_{p=1}^{2n} \mc{D}_p^{n_p} .
\end{equation}
Here, the set $\{ n_p \}$ is one way to distribute $2n$ copies of $W$ into $n_1$ numbers of one-body contractions, $n_2$ numbers of two-body contractions etc. and $\mathcal{S}\left(\{n_p\}\right)$ is the number of ways in which we can obtain that grouping. We sum over all such possible groupings of $2n$ copies of $W$. Thus the primed sum indicates the sum over $n_1$, $n_2$, $n_3$ etc. subject to the constraint: $\sum_{p=1}^{2n} p\;n_p=2n$. We can evaluate   
the first few $\mc{D}$'s for the purpose of later use.
\begin{eqnarray} \label{eq:D1D2}
    \mc{D}_1&=&\frac{1}{D} \, \text{Tr}\left( \rho_R \right) \notag \\
    \mc{D}_2&=&\frac{1}{D^2} \, \left( D\; \text{Tr}\left( \rho_R^2 \right)- \left(\text{Tr}\left( \rho_R \right)\right)^2-\frac{1}{D}\text{Tr}\left( \rho_R^2 \right) \right). \notag\\
\end{eqnarray}
The prefactors are the appropriate Weingarten factors in large $D$ limit. In computing the above, we have used the following fact about phase point operators (see \cite{Basu:2025uxw} for more details):
\begin{equation}
    \text{Tr}\left( A_R(\mathbf{u})^n \right)=
    \begin{cases}
        1 \;\;,\quad \cdots\;\;\; n\; \text{is odd}\\
        D \;\;, \quad \cdots \;\;n\; \text{is even}.
    \end{cases}
\end{equation}
From \eqref{eq:D1D2}, it follows that at large $D$, the two-body contribution $\mathcal{D}_2$ is approximately given by $\frac{1}{D^2}\;D\;\mathrm{Tr}(\rho_R^2)$. More generally, we get the following power counting:
\begin{equation} \label{eq:Dp}
    \mathcal{D}_p\approx
    \begin{cases}
        \frac{1}{D^p} \text{Tr}\left( \rho_R^p \right)\,\;\;\cdots \quad \quad p\; \text{is odd} \\
        \frac{1}{D^{p-1}}  \;\text{Tr}\left( \rho_R^p \right)\,\;\;\cdots  \quad p\; \text{is even}.
    \end{cases}
\end{equation}
%Note that the density matrix $\rho_R$ appearing here is the one that contains heavy operator insertions. Consequently, the scaling of quantities such as $\mathrm{Tr}(\rho_R^p)$ depends on the relative sizes of the left and right microcanonical windows. This dependence becomes manifest upon performing the average over the heavy operator matrix elements which we do next.
From equations \eqref{eq:W2n_haar} and \eqref{eq:Dp}, we see that the dominant contribution to $\langle W^{2n}\rangle$ in the large $D$ limit is given by the sum over all two-body contractions:\footnote{In taking the large $D$ limit, we do not worry about the invariants $\text{Tr}(\rho_R^n)$ etc., because the density matrix $\rho_R$ has support over microcanonical windows with fixed dimensions, which we can hold fixed as $D \to \infty$.}
\begin{equation} \label{eq:W2navgfinal}
    \langle W^{2n} \rangle_U \approx \frac{1}{\left( D \right)^{2n}}\times \frac{(2n)!}{2^n n!} \times \mc{D}_2^{n},
\end{equation}
where $\mc{D}_2$ is given by
\begin{equation} \label{eq:D2}
    \mathcal{D}_2 \approx \frac{1}{D}\, \text{Tr}\left( \rho_R^2\right),
\end{equation}
and the factor $\frac{(2n)!}{2^n n!}$ is a combinatorial factor which counts the number of ways of grouping $2n$ objects into pairs. Analytically continuing this formula to $n= \frac{1}{2}$ and summing over phase space, we get
\beq \label{eq:neg_rep_trick}
\mathcal{N}(\rho_R) = \sqrt{\frac{2}{\pi}} \left(\frac{D}{e^{S_2}}\right)^{1/2},
\eeq 
where $S_2(\rho_R)$ is the second R\'enyi entropy. In appendix \ref{appendix1}, we show how to reproduce the same formula from a more systematic approach, without resorting to the replica trick. 

\subsubsection{Variance}

Having computed the averaged Wigner negativity, we will now argue that fluctuations around this average are suppressed in the large $D$ limit. To quantify these fluctuations, we consider the variance, $\sigma^2=\langle \mathcal{N}^2 \rangle_U - \langle \mathcal{N} \rangle_U^2$. Using the same replica-trick reasoning as before, this quantity can be expressed as,
\begin{multline}
    \sigma^2=\sum_{\mathbf{u}}\sum_{\mathbf{v}}\lim_{m \to \frac{1}{2}} \lim_{n\to \frac{1}{2}}\Big[  \langle W^{2m}(\mathbf{u})\; W^{2n}(\mathbf{v}) \rangle_U -\\
    \langle  W^{2m}(\mathbf{u}) \rangle_U \; \langle  W^{2n}(\mathbf{v}) \rangle_U   \Big].
\end{multline}
One can then see that the only terms in $\langle W^{2m}(\mathbf{u}) W^{2n}(\mathbf{v}) \rangle_U$ that contribute to the variance are those involving contractions of $W$'s between the two replica sectors. More precisely, these are the contraction patterns in which phase point operators $A(\mathbf{u})$ and $A(\mathbf{v})$ contract with one another. The dominant contribution comes from contraction patterns containing $2,4,6,\ldots m+n$ pairs of cross-contractions of the type $\langle  W(\mathbf{u}) W(\mathbf{v})\rangle_U$, while all remaining $W$'s are contracted pairwise within their respective replica sectors. All such diagrams evaluate to the same value. Denoting by $\mathcal{S}(m,n)$ the total number of all such terms, we get,
\begin{multline}
    \sigma^2\approx \sum_{\mathbf{u}}\sum_{\mathbf{v}}\lim_{m \to \frac{1}{2}} \lim_{n\to \frac{1}{2}} \mathcal{S}(m,n) \; \frac{1}{D^{2m+2n}}\; \delta_{\mathbf{u}\mathbf{v}} \times \\
    \left( \frac{1}{D^2} \times D \; \text{Tr}\left( \rho_R^2 \right) \right)^{m+n}.
\end{multline}
Upon performing the analytic continuation and summing over the phase-space points, one finds that
$\sigma^2 \sim O\left(\frac{1}{D\;e^{S_2}}\right)$, and thus is vanishingly small as $D \to \infty$.

\subsection{Relative negativity}
So far, we have focused on computing the averaged Wigner negativity of the state $\rho_R$. One unappealing feature of equation \eqref{eq:neg_rep_trick} is that the formula depends on the dimension $D=d^n$ of the underlying microscopic Hilbert space. However, this is a universal factor that is present in every state supported on a small enough microcanonical window. We can get rid of this factor by defining the \emph{relative negativity}:
\beq 
\mathcal{N}(\rho_R | \rho_R^{(0)}) = \frac{\mathcal{N}(\rho_R)-1}{\mathcal{N}(\rho_R^{(0)})-1},
\eeq 
with $\rho^{(0)}_R$ being some reference state on $R$ restricted to the same microcanonical window $(E^\star_R-\epsilon, E^\star_R+\epsilon)$, where from the gravity point of view $E_R^{\star}$ is the ADM energy at the right asymptotic boundary for the PET state. Thus, using equation \eqref{eq:neg_rep_trick}, we get
\beq 
\mathcal{N}(\rho_R | \rho_R^{(0)}) \simeq e^{\frac{1}{2}\left(S_2(\rho_R^{(0)})-S_2(\rho_R)\right)}.
\eeq 
In the present content, it is natural to take $\rho^{(0)}_R$ to be the maximally mixed state in the energy window, i.e., the microcanonical ensemble. For $\rho_R^{(0)}$, we thus get
\beq 
S_2(\rho_R^{(0)}) = S_{\text{vN}}(\rho^{(0)}_R) \sim \frac{1}{4G_N}A(\gamma_{\text{out}}),
\eeq 
where $\gamma_{\text{out}}$ is the outermost extremal surface in the geometry dual to the PET state, and we have used the fact that in the microcanonical ensemble, the R\'enyi entropy equals the von Neumann entropy, which is given by the area of $\gamma_{\text{out}}$ at leading order. 

We need to also compute the second R\'enyi entropy for the original density matrix $\rho_R$ corresponding to the PET state. For this purpose, we use one further assumption:\\
\\
\noindent\textbf{Assumption 2}: \emph{The coefficients $R_{ij}$ appearing in the matrix elements $\langle E_i|\mathcal{O}_h|E_j\rangle$ of the heavy operator in energy eigenstates within the relevant microcanonical window are i.i.d Gaussian random variables with zero mean and unit variance.}
\\
\\
This makes the state $\Psi_\star$ in equation \eqref{eq:RTN} a random tensor network. For such states, averaging over $R_{ij}$ gives:
\beq
\mathrm{Tr}(\rho_R^2) = \frac{1}{e^{S(E^{\star}_L)}} + \frac{1}{e^{S(E^{\star}_R)}},
\eeq
and so
\begin{eqnarray}
S_2(\rho_R) &\simeq& \text{min}\Big(S(E^\star_L), S(E^\star_R)\Big)\nonumber\\
&\simeq &\frac{1}{4G_N}\text{min}\Big(A(\gamma_{\text{out}}), A(\gamma_{\text{in}})\Big),
\end{eqnarray}
where $\gamma_{\text{in}}$ is the inner extremal surface corresponding to the horizon of the left black hole.
Thus, the relative negativity is given by
%\beq \label{eq:diffSurf}
%\mathcal{N}(\rho_R|\rho^{(0)}_R)\simeq \exp\left[\frac{1}{8G_N}\left(A(\gamma_{\text{out}})- A(\gamma_{\text{in}})\right)\right].
%\eeq 
%We can equivalently write this formula as: 
\beq \label{eq:diffSurf2}
\mathcal{N}(\rho_R|\rho^{(0)}_R)\simeq \exp\left[\frac{1}{8G_N}\left(A(\gamma_{\text{out}})- A(\gamma_{\text{min}})\right)\right],
\eeq 
 where $\gamma_{\text{min}}$ is the extremal surface with minimal area. Note that the relative negativity is independent of the dimension of the microscopic Hilbert space, and only depends on the difference of the areas of the outer and inner surfaces at leading order in $\frac{1}{G_N}$. Thus, we learn that the stabilizer complexity gets exponentially enhanced when the bulk geometry contains a python's lunch region in the entanglement wedge of $R$. This is our main result.

\section{Discussion}\label{sec:disc}
In this note, we have studied the relative Wigner negativity between the reduced density matrix $\rho_R$ of the PET state with microcanonical boundary conditions and the maximally mixed state in the same energy window. With two simplifying assumptions pertaining to the chaotic nature of the Hamiltonian and the pseudorandom nature of the heavy operator insertion in the PET state, we have argued that the relative negativity shows an exponential enhancement when the bulk geometry has a python's lunch in the entanglement wedge of the corresponding boundary subregion $R$. The original python's lunch conjecture \cite{brown2020python} was about the complexity of bulk reconstruction, but here we have observed that a slightly different notion of complexity -- namely the stabilizer complexity of the boundary reduced density matrix -- gets exponentially enhanced in the presence of a python's lunch. Note that stabilizer complexity is usually defined with respect to a choice of computational basis, but for a microcanonical window of size $E_{\text{RMT}}$ where the spectrum displays random matrix universality (with $e^{-S(E^{\star})} \ll E_{\text{RMT}} \ll E^{\star}$), the answer turns out to be independent of the choice of computational basis, and thus purely information theoretic. 

While we have demonstrated the calculation explicitly in the case of the PET state with one heavy operator insertion, we expect the result in eqaution \eqref{eq:diffSurf2} to be true more generally for PET states with arbitrary number of operator insertions, but with fixed energy boundary conditions. The point is that equation \eqref{eq:neg_rep_trick} did not assume much about the density matrix $\rho$, and is more generally true as long as assumption 1 is satisfied. Furthermore, with fixed energy boundary conditions and when assumption 2 is satisfied for all heavy operator insertions, then the boundary state $\Psi_\star$ turns into a random tensor network, for which the second R\'enyi entropy is generally given by minimizing the generalized entropy over all choices of extremal surfaces. The second assumption is also true of \emph{fixed area states} in holography \cite{Dong:2018seb, Akers:2018fow}, so we expect our result to hold for these states as well. We also expect our result to apply to multi-boundary black hole states in $AdS_3$ \cite{Balasubramanian:2014hda} projected onto microcanonical energy windows, provided we make appropriate pesudorandomness assumptions on the OPE coefficients \cite{belin2021random, Chandra:2022bqq}. 

We end with two cautionary remarks: firstly, equation \eqref{eq:diffSurf2} may \emph{not} apply more generally to holographic states with Dirichlet (as opposed to fixed-energy) boundary conditions. It would be interesting to pursue this generalization. However, we do not view this as a problem -- the point of this paper was to interpret the geometric notion of complexity associated to a python's lunch from a boundary CFT point of view, and indeed, we have argued that the python's lunch finds a natural interpretation in terms of the stabilizer complexity of the reduced density matrix corresponding to the microcanonical state. Secondly, in our setup, the region of spacetime outside the outer extremal surface is identical to a black hole with the correct ADM energy, with no additional excitations. One could more generally consider a sitation where there is additional matter in this region, resulting in the causal wedge of $R$ being smaller than the outer wedge of $R$. It has been argued in \cite{Engelhardt:2021mue} that by the action of simple unitaries on the region $R$ the entire outer wedge can be brought into the causal wedge, and the effect of such matter perturbations removed. One would expect this operation to have $O(1)$ complexity, provided one is allowed access to an $O(1)$ (not scaling with $N$) number of applications of the boundary Hamiltonian and simple (i.e., single trace) operators to the CFT subregion $R$. So far we have not said anything specific about the choice of computational basis on $R$, but the requirement that we assign an $O(1)$ stabilizer complexity to quantum operations involving an $O(1)$ number of applications of single trace operators and the Hamiltonian should constrain the choice of computational basis on $R$ to some extent. If such a choice of basis is possible,\footnote{A version of the Krylov basis construction of \cite{Basu:2024tgg, Basu:2025mmm, Balasubramanian:2026klv, Basu:2026aky} could be useful in finding such a basis.} then it would perhaps also alleviate our first concern above to some extent, because we could project any given state to an $O(\frac{1}{\log N})$ microcanonical window on the boundary by an operation of $O(\log N)$ complexity. 

\subsection*{Acknowledgements}
We would like to thank Chris Akers, Sriram Akella, Abhijit Gadde, Shiraz Minwalla, Pratik Rath, Ronak Soni and Sandip Trivedi for helpful discussions and comments. We are particularly grateful to Pratik Rath for several illuminating conversations on the Thouless energy and on the python's lunch conjecture. OP acknowledges fruitful discussions during the workshop ``Observers, wormholes and complex saddles in cosmology", organized at the Bernoulli Center for Fundamental Studies (EPFL, Lausanne) from 18--22 May 2026. Reseach supported by the Department of Atomic Energy, Government of India, under Project Identification Number RTI-4012 and from the Infosys Endowment for the study of the Quantum Structure of Spacetime.

\appendix

\section{Integral representation method}\label{appendix1}

So far, we have employed a naive replica-trick approach to compute the average of the absolute value of the Wigner function, $\langle |W|  \rangle_U$. A more systematic treatment can be carried out using the integral representation:
\begin{equation}
    |W(\mathbf{u} )|=\lim_{\epsilon \to 0}\frac{1}{2\pi i} \int_{-\infty}^{\infty} \;dz \frac{2z}{z^2+\epsilon^2} \; W(\mathbf{u} )\; e^{izW(\mathbf{u} )}
\end{equation}
We are then supposed to compute the average $\langle W e^{izW} \rangle_U$ which in turn can be computed as, $-i \frac{\partial}{\partial z} \langle e^{izW} \rangle_U$. Hence the negativity is given by,
\begin{equation} \label{eq:Neg_int}
    \mathcal{N}= \sum_{\mathbf{u}}\lim_{\epsilon \to 0}\frac{1}{2\pi i} \int_{-\infty}^{\infty} \;dz \frac{2z}{z^2+\epsilon^2} \; \left(-i \frac{\partial}{\partial z} \right) \langle e^{izW( \mathbf{u})} \rangle_U
\end{equation}
To obtain an expression for $\langle e^{izW}\rangle_U$, one essentially expands the exponential and computes the quantities $\langle W^{n} \rangle_U$ according to the formula \eqref{eq:W2n_haar}. The symmetry factor is
$\mathcal{S}(\{n_p\})=\frac{n!}{\prod_{p=1}^{n} (p!)^{n_p}\, n_p!},$ and hence
\begin{multline}
    \langle W^{n}\rangle_U=\frac{1}{\left( D \right)^{n}}\sum_{\{n_p \}}'  \frac{n!}{(1!)^{n_1}(2!)^{n_2}\cdots (n!)^{n_n}} \times \\
    \frac{\mc{D}_1^{n_1}\mc{D}_2^{n_2}\cdots \mc{D}_n^{n_n}}{n_1!n_2!\cdots n_n!}  .
\end{multline}
Of course, there is no closed-form expression for this sum. However, we will
eventually compute $\sum_{n=0}^{\infty}\frac{(iz)^n}{n!}\langle W^n\rangle_U$.
This resummation can be carried out exactly (see \cite{Basu:2025uxw} for details), yielding
\begin{equation}
    \langle e^{izW}\rangle_U = \exp \left[ \sum_{p=1}^{\infty} \left( \frac{iz}{D} \right)^p \frac{\mc{D}_p}{p!}\right]
\end{equation}
$\mathcal{D}_p$ were defined in the main text. We will again retain only the leading-order contribution to each $\mathcal{D}_p$, \eqref{eq:Dp}. Thus we find,
\begin{eqnarray}
    \langle  e^{izW}\rangle_U &\approx& \exp \Bigg[ \sum_{p=1}^{\infty} \left( \frac{iz}{D^2} \right)^{2p-1} \frac{\text{Tr}\left( \rho_R^{2p-1} \right)}{(2p-1)!} \notag \\
    &+& \sum_{p=1}^{\infty} \left( \frac{iz}{D^2} \right)^{2p} \frac{D\;\text{Tr}\left( \rho_R^{2p} \right)}{(2p)!} \Bigg]
\end{eqnarray}
We may now substitute this expression into \eqref{eq:Neg_int} to compute the Wigner negativity. Recall that, $\mathrm{Tr}(\rho_R^p)=e^{\left( 1-p \right)S_p(\rho_R)}$ which roughly scales as $e^{(1-p)S}$. At this stage, it is convenient to introduce the rescaled variable $t = \frac{z}{De^{S_2}}$. Using these facts, one finds that the exponent takes the form,
\begin{multline}
    \exp \left[\frac{e^{S_2}}{D} \left(it -\frac{t^2}{2}-\frac{1}{D^2} \frac{it^3}{3!}\frac{\left( e^{S_2} \right)^2}{e^{2S_3}}+ 
    \frac{1}{D^3}\frac{t^4}{4!}\frac{\left( e^{S_2} \right)^3}{e^{3S_4}}+\cdots \right)\right] \\
\end{multline}
Following the arguments presented in \cite{Basu:2025uxw}, one can keep only the first two terms in the exponent. Then we have,
\begin{multline}
    \mathcal{N}\approx \sum_{\mathbf{u}}\frac{1}{D \;e^{S_2}}\lim_{\epsilon \to 0}\frac{1}{2\pi i} \int_{-\infty}^{\infty} \;dt \frac{2t}{t^2+\epsilon'^2} \;\times \\
    \left(-i \frac{\partial}{\partial t} \right)\; \exp \left[ \frac{e^{S_2}}{D} \left(it -\frac{t^2}{2}\right)\right]
\end{multline}
The integral over $t$ can be carried out exactly. Note that the dependence on $\mathbf{u}$ disappears after performing the Haar average. Consequently, the phase-space sum becomes trivial, and one ultimately obtains the following expression for the Wigner negativity:
\begin{equation}
    \mathcal{N}=\text{erf} \left( \sqrt{r}\right)+\frac{e^{-r}}{\sqrt{\pi r}} \quad, \quad r=\frac{e^{S_2}}{2D}
\end{equation}
In our present discussion $D \gg e^{S_2}$, and as a result, one can show that the above expression reduces to,
\begin{equation}
    \mc{N}=\sqrt{\frac{2}{\pi}}\times \sqrt{\frac{D}{e^{S_2}}}
\end{equation}
which is the same formula we had obtained in \eqref{eq:neg_rep_trick}, using the replica trick.

\section{Test of Assumption-1 in the SYK model}\label{appendix2}

\begin{figure*}[t]
    \centering
    \includegraphics[width=\textwidth]{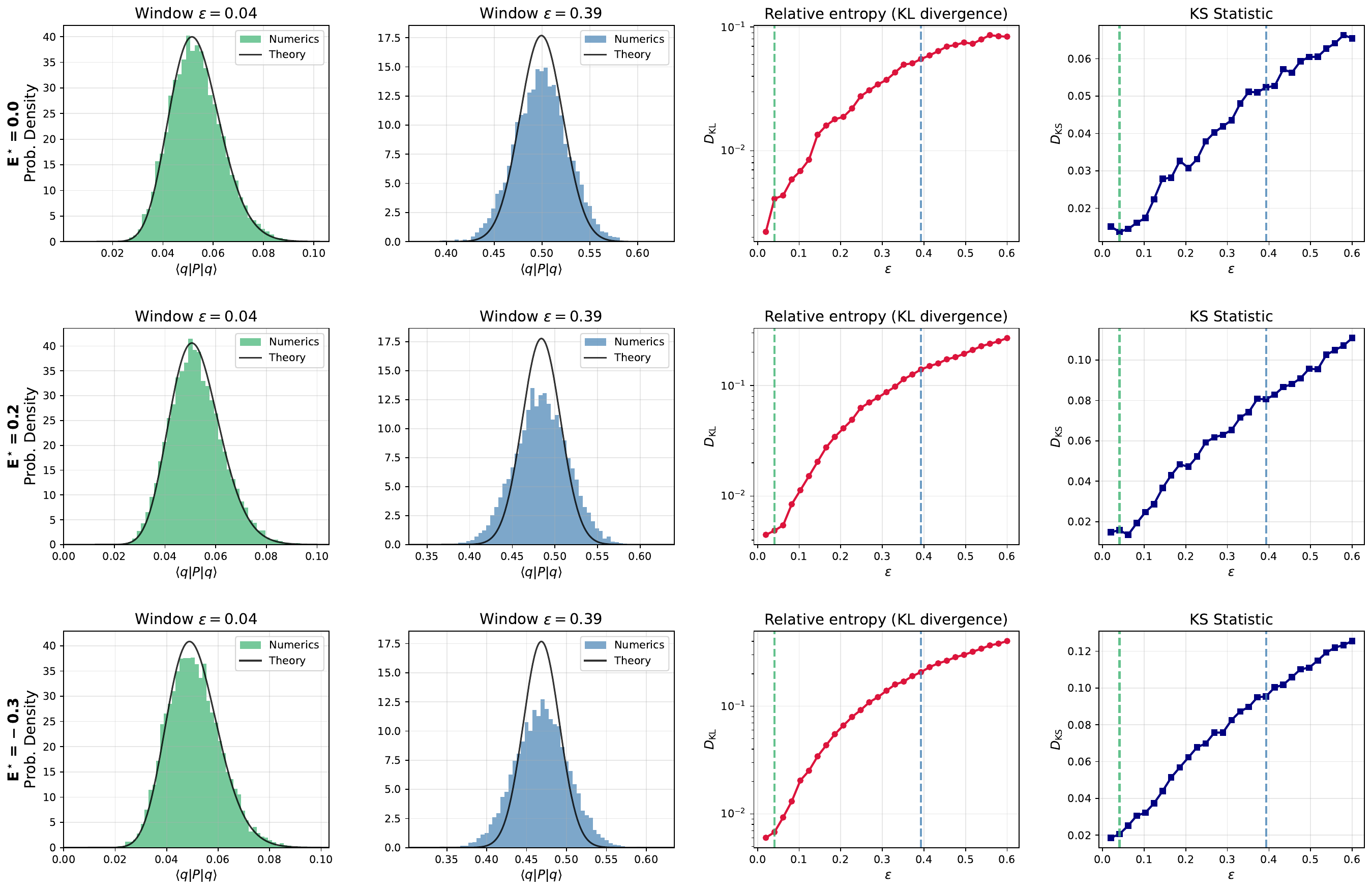}
    \caption{Numerical evidence for Assumption-1 in the SYK model with $N=20$, averaged over 20 realizations. Rows correspond to three different energy centers ($E^\star = 0.0, 0.2, -0.3$). The first two columns compare the numerical distribution of $\langle \bq | P | \bq\rangle$ against the exact theoretical prediction for two representative microcanonical window sizes ($\epsilon=0.04$ and $\epsilon=0.39$). First column: for sufficiently small values of $\epsilon$, our assumption holds
very well. Second column: as $\epsilon$ increases beyond the RMT energy scale,
$\epsilon > E_{\mathrm{RMT}}$, the assumption begins to break down. The rightmost columns quantify the degree of validity of this randomness assumption, plotting the KL and KS divergences as a function of the increasing window half-width $\epsilon$.}
    \label{fig:syk_numerics1}
\end{figure*}

In this Appendix, we provide preliminary evidence in support of Assumption-1 in a particular model, namely the Sachdev-Ye-Kitaev (SYK) model \cite{Sachdev_1993, Kitaev1, Kitaev2, Maldacena:2016hyu}. We consider the SYK model consisting of $N$ Majorana fermions. The Hamiltonian is given by
\begin{equation}
    H=\sum_{1\le i_1<i_2<i_3<i_4\le N} \; J_{i_1 i_2 i_3 i_4} \; \psi_{i_1} \psi_{i_2} \psi_{i_3} \psi_{i_4},
\end{equation}
where $\psi_i$ are Majorana fermions and the coefficients $J_{i_1i_2i_3i_4}$ are drawn from a random gaussian distribution with zero mean and variance given by $\langle J_{i_1 i_2 i_3 i_4}^2 \rangle=\frac{6J^2}{N^3}$. In our analysis, we will set $J=1$, so all energies will be measured in these units.

Given this Hamiltonian, we can find its spectrum and choose a reference
energy $E^\star$. We then define the corresponding microcanonical subspace as
$\mathcal{H}(E^\star)=\operatorname{span}\left\{\ket{E_i}\,\middle|\,E^\star-\epsilon < E_i < E^\star+\epsilon\right\}$.
According to Assumption-1, for sufficiently small $\epsilon$, this subspace behaves like a random subspace relative to the computational basis. In order to test this assumption, we proceed as follows: let us consider the projector onto the microcanonical subspace, $P=\sum_{i=1}^{k}\ket{E_i}\bra{E_i}$, where
$k$ is the dimension of the microcanonical subspace. We then compute the quantity $x=\langle \bq|P|\bq\rangle$, where $\ket{\bq}$ denotes a computational basis
state:
\begin{equation}
    x=\sum_{i=1}^{k} |\langle \bq |E_i\rangle|^2=\sum_{i=1}^{k} |U_{\bq,i} |^2.
\end{equation}

\noindent If we now use our assumption \eqref{eq:MBBC} and take $U_{q,i}$ to be elements of a Haar-random unitary, then it is possible to show that the quantity $x$ must follow the Beta distribution \cite{Zyczkowski_2000}:
\begin{equation}
    P_{\mathrm{theo}}(x) = \frac{\Gamma\left(\frac{D}{2}\right)}{\Gamma\left(\frac{k}{2}\right)\Gamma\left(\frac{D-k}{2}\right)} x^{\frac{k}{2} - 1} (1 - x)^{\frac{D-k}{2} - 1},
    \label{eq:theoretical_distribution}
\end{equation}
where $D= 2^{N/2}$ is the dimension of the full Hilbert space. So, we can test our assumption by computing $x$ for several different computational basis states and verifying the above expectation.

In figure \ref{fig:syk_numerics1}, we present our preliminary findings. For a fixed
choice of $E^\star$, we consider several values of $\epsilon$ and construct the
corresponding projector $P$. Here, $\ket{\bq}$ denotes a computational basis state
in the qubit representation of the Hilbert space. We then compute $\langle \bq|P|\bq\rangle$ numerically,
plot its histogram, and compare it with the theoretical prediction. We find
that, for small enough values of $\epsilon$ (see the first column of Fig \ref{fig:syk_numerics1}), the numerical results agree well with the theoretical
prediction. Furthermore, to quantify the deviation of the numerical distribution from the theoretical prediction $P_{\text{theo}}$, we employ two complementary statistical measures. The Kullback-Leibler (KL) divergence evaluates the information-theoretic relative entropy between the binned numerical probability distribution $P_{\mathrm{num}}$ and the exact theoretical prediction $P_{\mathrm{theo}}$, defined over discrete bins $i$ as 
\begin{equation}
    D_{\mathrm{KL}} = \sum_{i} P_{\mathrm{num}, i} \ln \left( \frac{P_{\mathrm{num}, i}}{P_{\mathrm{theo}, i}} \right).
\end{equation}
To provide a stringent, un-binned evaluation, we also compute the Kolmogorov-Smirnov (KS) statistic, which captures the maximal absolute distance between the respective cumulative distribution functions (CDFs), given by 
\begin{equation}
    D_{\mathrm{KS}} = \sup_{x} \left| \mathrm{CDF}_{\mathrm{num}}(x) - \mathrm{CDF}_{\mathrm{theo}}(x) \right|.
\end{equation}
We plot these quantities for different sizes of the microcanonical window; see
the third and fourth columns of Fig. \ref{fig:syk_numerics1}. For $N=20$, the expected RMT energy scale
$E_{\mathrm{RMT}}\sim 1/\log N$ is approximately of the order of $0.3$. We find good agreement
between the numerical results and the theoretical expectation for a random
subspace for values of $\epsilon$ smaller than $0.3$.

\bibliographystyle{apsrev4-2}
\bibliography{references}

\end{document}